\documentstyle[prl,aps,float,epsfig]{revtex}

\begin{document}
\draft
\wideabs{
\title{Implementation of the refined Deutsch-Jozsa algorithm on a 3-bit NMR quantum computer}
\author{Jaehyun~Kim, Jae-Seung~Lee, and Soonchil~Lee}
\address{Department of Physics, Korea Advanced Institute of Science %
and Technology, Taejon, 305-701, Korea}
\author{Chaejoon Cheong}
\address{Magnetic Resonance Team, Korea Basic Science Institute, Taejon, 305-333, Korea}
\date{\today}
\maketitle
\begin{abstract}
We implemented the refined Deutsch-Jozsa algorithm on a 3-bit nuclear magnetic
resonance quantum computer, which is the meaningful test of quantum parallelism
because qubits are entangled. All of the balanced and constant functions were
realized exactly. The results agree well with theoretical predictions and clearly
distinguish the balanced functions from constant functions. Efficient refocusing
schemes were proposed for the soft $z$-pulse and J-coupling and it is proved that
the thermal equilibrium state gives the same results as the pure state for this
algorithm.
\end{abstract}

\pacs{PACS numbers : 03.67.Lx, 03.67.-a, 76.90.+d} }
A quantum computer which was just a theoretical concept has been realized recently
by nuclear magnetic resonance (NMR). Several methods have been proposed such as ion
trap~\cite{cirac,monroe}, quantum dot~\cite{barenco,loss}, cavity
QED~\cite{turchette,domokoss}, and Si-based nuclear spins~\cite{kane} to realize
quantum computers but NMR~\cite{gershenfeld} has given the most successful results.
Several quantum algorithms have been implemented by NMR quantum
computers\cite{chuang1,chuang2,jones1,jones2,linden1} among which the Deutsch-Jozsa
(D-J) algorithm~\cite{d-j} has been studied most because it is the simplest quantum
algorithm that shows the power of a quantum computer over a classical one. Most of
quantum algorithms, including the D-J algorithm, have been implemented only for
functions of one and two bits. The successful implementation of a quantum algorithm
depends heavily on the number of basic operations which increases with the number
of qubits due to finite coherence time. Moreover, more than 2-bit operations
require more than two-body interactions which do not exist in nature. It is
possible to avoid such interactions, though not easy, but it increases again the
number of total basic gates and coherence may break down during the computation.
There have been few works that have performed real three-bit
operations~\cite{weinstein} so far.

The D-J algorithm determines whether an $n$-bit binary function,
\begin{equation}
f : \{0,1\}^n \longmapsto \{0,1\},
\end{equation}
is a constant function which always gives the same output, or a balanced function
which gives 0 for half of inputs and 1 for the remaining half. The D-J algorithm
gives answer by only one evaluation of the function while a classical algorithm
requires $(2^{n-1}+1)$ evaluations in the worst case. The function is realized in
quantum computation by unitary operation,
\begin{equation}
U |x\rangle |y\rangle = |x\rangle |y\oplus f(x)\rangle,
\end{equation}
where $x$ is an $n$-bit argument of the function and $y$ is one-bit. If $|y\rangle$
is in the superposed state, $(|0\rangle -|1\rangle)/\sqrt2$, then the result of the
operation,
\begin{equation}
U |x\rangle (\frac{|0\rangle -|1\rangle}{\sqrt2})
 = (-1)^{f(x)} |x\rangle (\frac{|0\rangle -|1\rangle}{\sqrt2}),
\label{eq-dj}
\end{equation}
carries information about the function encoded in the overall phase. If $|x\rangle$
is also prepared in the superposition of all its possible states, $(|0\rangle
+|1\rangle + \cdots |2^n-1\rangle )/\sqrt{2^n}$, by applying an $n$-bit Hadamard
operator $H$ to $|x\rangle=|0\rangle$, the relative phases of the $2^n$ states
change depending on $f$. If $f$ is a constant function, then the relative phases
are all same and additional application of $H$ restores $|x\rangle$ to $|0\rangle$.
If $f$ is a balanced function, $|x\rangle$ cannot be restored to $|0\rangle$ by
this operation. It is obvious that $|y\rangle$, being in the superposed state,
$(|0\rangle -|1\rangle)\sqrt2$, plays a central role in the algorithm but it is
redundant in the sense that its state does not change.

This redundancy is removed in the refined D-J algorithm~\cite{collins} where the
following unitary operator is used.
\begin{equation}
U_f |x\rangle = (-1)^{f(x)} |x\rangle.
\end{equation}
It has been shown that $U_f$ is always reduced to a direct product of single-bit
operators for $n \leq 2$. In this case, $n$ classical computers can do the same job
by simultaneous evaluations because qubits are never entangled. Therefore,
meaningful tests of the D-J algorithm can occur if and only if $n>2$. Recently, a
realization of the D-J algorithm for $n=4$ has been reported~\cite{marx}, but in
that work, only one balanced function was evaluated and the corresponding $U_f$ is
reducible to a direct product of four single-bit operators. In this study, we
investigated the refined D-J algorithm with 3-bit arguments to find out the pulse
sequences of $U_f$'s, and implemented the algorithm on an NMR quantum computer for
all the functions.

There are $_8\text{C}_4=70$ balanced and two constant functions among all 3-bit
binary functions. We index the functions with their outputs, $f(0)\cdots f(7)$,
expressed as hexadecimal numbers. For example, $f_{{\tt 1E}}$ denotes the function
of which the outputs are given by $f(0)\cdots f(7)=00011110$. Note that $U_{f_{{\tt
x}}}=-U_{f_{{\tt FF}-{\tt x}}}$, where {\tt x} is a hexadecimal number equal to or
less than {\tt FF}. The difference of overall phase cannot be distinguished in the
experimental implementations. Therefore, there are 35 distinct unitary operators
corresponding to the balanced functions, and one operator corresponding to the
constant functions. Since the unitary operator corresponding to the constant
functions, $U_{f_{{\tt 00}}}$, is just the unity matrix, there are 35 non-trivial
and distinct $U_f$'s to be implemented.

The NMR Hamiltonian of the weakly interacting three spin system is given by
\begin{equation}
{\cal H}= \sum^3_i \Delta\omega_i I_{iz} +\sum^3_{i<j} \pi J_{ij}2I_{iz}I_{jz}
\label{eq-hamil}
\end{equation}
in the rotating frame, where $I_{iz}$ is the $z$-component of the angular momentum
operator of spin $i$. The first term represents the precession of spin $i$ about
$z$-axis due to the chemical shift, $\Delta\omega_i$, and the second term the
spin-spin interaction between spin $i$ and $j$ with coupling constant $J_{ij}$.
This Hamiltonian provides six unitary operators, $I_{iz}(\theta)=\exp[-\imath\theta
I_{iz}]$ and $J_{ij}(\theta)=\exp[-\imath\theta 2I_{iz}I_{jz}]$. In combination
with $I_{iz}(\theta)$, two other operators $I_{ix}(\theta)$ and $I_{iy}(\theta)$
produced by rf pulses can perform any single-bit operations. The coupling operator
$J_{ij}(\theta)$ can be used to make a controlled-{\sc not} operation. The
combination of single-bit operations and controlled-{\sc not} operations can
generate any unitary operations~\cite{barenco2}.

Table~\ref{tbl-35} shows the sequences of the realizable operators for all the 35
non-trivial distinct $U_f$'s. In the table, the notations $I_1$, $I_2$ and $I_3$
were replaced by $I$, $S$ and $R$, respectively for convenience. Some of $U_f$'s
are irreducible and require three-body interaction. The sequences of realizable
operators in the table were obtained by following a general implementation
procedure using generator expansion~\cite{jhkim}. This method includes the coupling
order reduction technique which replaces an $n$-body interaction operator for $n>2$
by two-body ones. It is noticed that all $U_f$'s consist of the operators of the
single-spin rotations about $z$-axis and spin-spin interactions only. From now on,
we call pulses corresponding to these operators the soft $z$-pulse and J-coupling,
respectively.

The balanced functions are classified into four types depending on the number of
$J_{ij}(\theta)$'s included in their operation sequences. It is easy to see that no
qubits are entangled in type-I functions and therefore, obviously they are not the
cases of meaningful tests. In type-II functions, only two qubits out of three are
entangled. So, type-II functions can be said to be the stepping stones to
meaningful tests. In type-III and IV functions, all three qubits are entangled and
the functions of these types can be tested only by a three-bit quantum computer.
Therefore, the realization of type-III and IV functions demonstrates true quantum
parallelism. Each sequence in Table~\ref{tbl-35} is not unique for a given function
but we believe that they are optimal ones for implementation of the refined D-J
algorithm.

The whole operation sequence for implementation of the refined D-J algorithm is
given by $H$-$U_f$-$H$-$D$ to be read from left to right. The first and second
$H$'s were realized by hard $\pi/2$ and $-\pi/2$ pulses about $y$-axis,
respectively. Since the read-out operation $D$ can be realized by a hard $\pi/2$
pulse about $y$-axis, the second $H$ and $D$ cancel each other to make the sequence
$H$-$U_f$.

The superposed input state is generated by the Hadamard operation on the pure state
$|0\rangle$. Therefore, it is usually necessary to convert the thermally
equilibrated spin state into the effective pure state. In the case of the refined
D-J algorithm, however, the thermal equilibrium state gives the same results with
the pure state. The deviation density matrix of the thermal equilibrium state,
$\rho_{\text{th}}$, is approximated by
\begin{equation}
\rho_{\text{th}}=I_{1z}+I_{2z}+I_{3z}
\end{equation}
for the Hamiltonian of Eq.~\ref{eq-hamil}, and the density matrix of $|0\rangle$,
$\rho_{\text{p}}$, is given by
\begin{equation}
\begin{array}{rcl}
\rho_{\text{p}} &=& I_{1z}+I_{2z}+I_{3z} \\
&&\mbox{}+2I_{1z}I_{2z}+2I_{2z}I_{3z}+2I_{1z}I_{3z}+4I_{1z}I_{2z}I_{3z} \\ &=&
\rho_{\text{th}} + \Delta\rho.
\end{array}
\end{equation}
The hard $\pi/2$ pulse for $H$ transforms terms of $\rho_{\text{th}}$ into
single-quantum coherence and terms of $\Delta\rho$ into multiple-quantum
coherence~\cite{ernst}. Since the sequences for $U_f$'s consist of only the soft
$z$-pulse(s) and J-coupling(s) which are dependent only on the $z$-components of
spin angular momentums, $U_f$'s do not change the order of quantum coherence. As
single-quantum coherence is only observable, $\rho_{\text{th}}$ and
$\rho_{\text{p}}$ give the same results for this case. In general, the thermal
equilibrium state gives the same results with the pure state if the operation
sequence after the first Hadamard operator does not change the order of quantum
coherence.

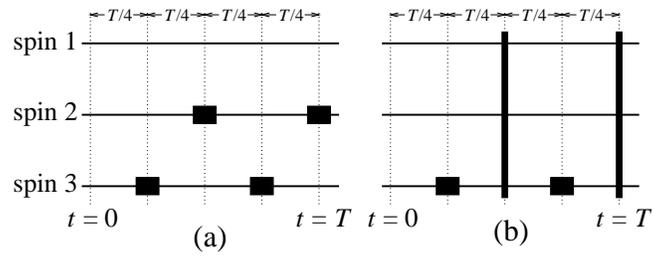
\begin{figure}
\centering
\caption{Refocusing schemes for (a) $I_z(\theta)$ and (b) $J_{12}(\theta)$. %
Short and long bars represent soft and hard $\pi$-pulses, respectively. %
The angle $\theta$ can be changed by adjusting the length of evolution time, $T$.}
\label{fig-zj}
\end{figure}

The soft $z$-pulse and J-coupling were implemented by the time evolution under the
Hamiltonian of Eq.~\ref{eq-hamil} with refocusing $\pi$-pulses applied at suitable
times during the evolution period. Since the refocusing $\pi$-pulse has the effect
of time reversal, it can be used to make one term in the Hamiltonian evolve while
the other terms ``freeze''~\cite{linden2,leung,jones3}. We optimized this {\it
refocusing scheme} as illustrated in Fig.~\ref{fig-zj} which shows the  soft
$z$-pulse on spin 1 and J-coupling between spin 1 and 2 as examples. The evolution
time, $T$, is $\theta/\Delta\omega_i$ for the soft $z$-pulse and $\theta/(\pi
J_{ij})$ for the J-coupling. Previous schemes divide the evolution period into
eight periods and require six pulses, or suffer from TSETSE
effect~\cite{linden3,kupce} because soft pulses exciting more than one but not all
spins were used. Since the difficulty of experiment increases exponentially with
increasing number of pulses, especially soft pulses, our scheme greatly enhances
the possibility of successful implementation. Axes of successive $\pi$-pulses were
chosen in the way to cancel imperfections of pulses. For example, four $\pi$-pulses
in Fig.~\ref{fig-zj}(a) were applied along $x$, $-x$, $-x$, and $x$-axes,
respectively.

In our experiment, $^{13}$C nuclear spins of 99\% carbon-13 labeled alanine
(CH$_3$CH(NH$_2$)CO$_2$H) in D$_2$O solvent were used as qubits. NMR signals were
measured by using a Bruker DRX300 spectrometer. The chemical shifts of three
different carbon spins are about 5670, $-3780$, and $-6380$~Hz, and coupling
constants $J_{12}$, $J_{23}$, and $J_{13}$ are 54.06, 34.86, and 1.03 Hz,
respectively. Protons were decoupled during the whole experiments. Gaussian shaped
soft $\pi$-pulses were 2~ms in length and hard pulses were about a few microsecond.
The length of the total pulse sequence was about 600~ms in the worst case.

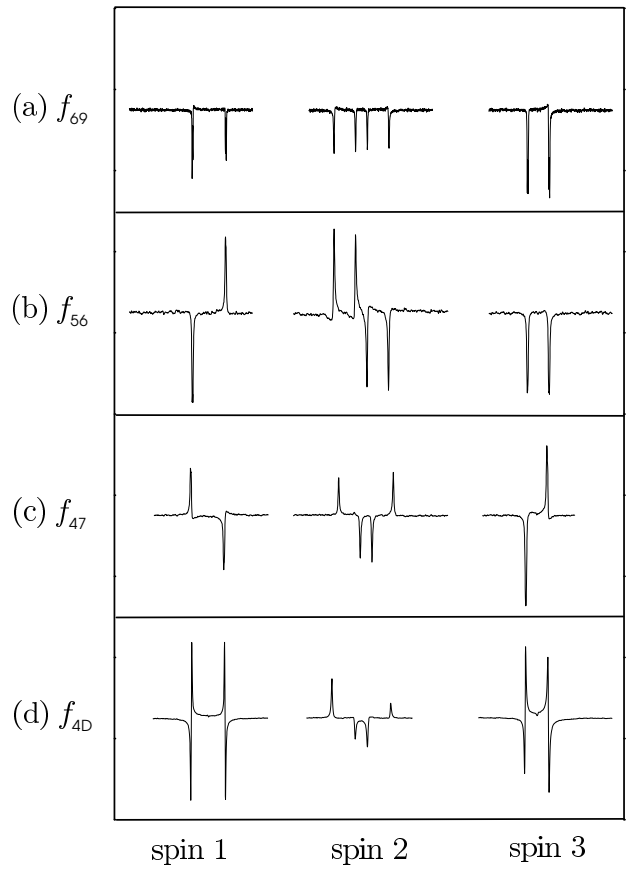
\begin{figure}
\centering
\caption{Observed signals for (a) $f_{{\tt 69}}$, (b) $f_{{\tt 56}}$, %
(c) $f_{{\tt 47}}$, and (d) $f_{{\tt 4D}}$. The $x$-axis represents frequency
increasing from right to left.} \label{fig-sig}
\end{figure}

We implemented all the 35 balanced and one constant functions exactly.
Fig~\ref{fig-sig} shows the results for the four functions belonging to different
types shown in Table~\ref{tbl-35}. The lines of the spectra for the remaining
functions also indicate as clearly as ones in the figure whether they are positive
or negative. The balanced functions are distinguished from the constant function
because some of the lines are negative. The peaks of spin 1 and 3 show up as
doublets in Fig~\ref{fig-sig}(a), (b) and (c) while that of spin 2 is quartet
because $J_{13}$ is very small compared to $J_{12}$ and $J_{23}$.
Fig~\ref{fig-sig}(d) shows, however, that the peaks of spin 1 and 3 are in fact
quartet also. They look dispersive doublets because the neighboring lines split a
little by $J_{13}$ have different signs. These results agree well with the
theoretical predictions obtained from
\begin{equation}
\text{Tr}(e^{\imath{\cal H}t/\hbar}\rho\;e^{-\imath{\cal H}t/\hbar} I_+),
\end{equation}
where $\rho$ is the density matrix transformed by the operation sequence $H$-$U_f$
from $\rho_{\text{th}}$ and $I_+=I_x+\imath I_y$.

In the implementation of the soft $z$-pulses and J-couplings shown in
Fig.~\ref{fig-zj}, the end of pulse sequence ($t=T$) can be clearly defined for the
J-coupling but not for the soft $z$-pulse, because the last pulse of soft $z$-pulse
is a soft pulse which is much longer than a hard pulse. Therefore, whole pulse
sequence was arranged to finish with the J-coupling. Our refocusing scheme
decreases the length of the total pulse sequence and therefore, reduces signal
decay due to decoherence. Imperfection of soft pulses is thought to be the main
source of the phase error and the decay of signal amplitude of some lines. This
imperfection is more serious in the J-coupling than in the soft $z$-pulse because
out-of-phase multiplets are produced in the former while in-phase multiplets are
produced in the latter. Therefore, it is very important to calibrate soft pulses
exactly especially for long sequences.

In summary, we implemented the complete refined D-J algorithm with 3-bit arguments
which involves entanglement. All the operations were realized by the time evolution
under Hamiltonian with refocusing $\pi$-pulses. The operation sequences best for
our implementation were found using generator expansion. Experimental pulse
sequences were made as simple as possible by using the thermal equilibrium state
and the new refocusing scheme.

\begin{table}
\widetext
\centering
\caption{Sequences of realizable operators for $U_f$'s corresponding to %
35 balanced functions.} \label{tbl-35}
\begin{tabular}{p{2em}c|p{2em}c|p{2em}c}
\multicolumn{2}{c|}{Type-I} &
$f_{{\tt 36}}$ & $S_z(\pi)I_z(-\frac{\pi}{2})R_z(-\frac{\pi}{2})J_{13}(\frac{\pi}{2})$ &
$f_{{\tt 3A}}$ & $S_z(\frac{\pi}{2})R_z(-\frac{\pi}{2})J_{12}(\frac{\pi}{2})J_{13}(\frac{\pi}{2})$ \\ \hline
$f_{{\tt 0F}}$ & $I_z(\pi)$ &
$f_{{\tt 39}}$ & $S_z(\pi)I_z(\frac{\pi}{2})R_z(-\frac{\pi}{2})J_{13}(\frac{\pi}{2})$ &
$f_{{\tt 53}}$ & $S_z(\frac{\pi}{2})R_z(\frac{\pi}{2})J_{12}(-\frac{\pi}{2})J_{13}(\frac{\pi}{2})$ \\ \hline
$f_{{\tt 33}}$ & $S_z(\pi)$ &
$f_{{\tt 63}}$ & $S_z(\pi)I_z(-\frac{\pi}{2})R_z(\frac{\pi}{2})J_{13}(\frac{\pi}{2})$ &
$f_{{\tt 35}}$ & $S_z(\frac{\pi}{2})R_z(\frac{\pi}{2})J_{12}(\frac{\pi}{2})J_{13}(-\frac{\pi}{2})$ \\ \hline
$f_{{\tt 55}}$ & $R_z(\pi)$ &
$f_{{\tt 6C}}$ & $S_z(\pi)I_z(\frac{\pi}{2})R_z(\frac{\pi}{2})J_{13}(\frac{\pi}{2})$ &
$f_{{\tt 5C}}$ & $S_z(-\frac{\pi}{2})R_z(\frac{\pi}{2})J_{12}(\frac{\pi}{2})J_{13}(\frac{\pi}{2})$ \\ \hline
$f_{{\tt 3C}}$ & $I_z(\pi)S_z(\pi)$ &
$f_{{\tt 56}}$ & $R_z(\pi)I_z(-\frac{\pi}{2})S_z(-\frac{\pi}{2})J_{12}(\frac{\pi}{2})$ &
$f_{{\tt 2E}}$ & $I_z(\frac{\pi}{2})R_z(-\frac{\pi}{2})J_{12}(\frac{\pi}{2})J_{23}(\frac{\pi}{2})$ \\ \hline
$f_{{\tt 66}}$ & $S_z(\pi)R_z(\pi)$ &
$f_{{\tt 59}}$ & $R_z(\pi)I_z(\frac{\pi}{2})S_z(-\frac{\pi}{2})J_{12}(\frac{\pi}{2})$ &
$f_{{\tt 47}}$ & $I_z(\frac{\pi}{2})R_z(\frac{\pi}{2})J_{12}(-\frac{\pi}{2})J_{23}(\frac{\pi}{2})$ \\ \hline
$f_{{\tt 5A}}$ & $R_z(\pi)I_z(\pi)$ &
$f_{{\tt 65}}$ & $R_z(\pi)I_z(-\frac{\pi}{2})S_z(\frac{\pi}{2})J_{12}(\frac{\pi}{2})$ &
$f_{{\tt 1D}}$ & $I_z(\frac{\pi}{2})R_z(\frac{\pi}{2})J_{12}(\frac{\pi}{2})J_{23}(-\frac{\pi}{2})$ \\ \hline
$f_{{\tt 69}}$ & $I_z(\pi)S_z(\pi)R_z(\pi)$ &
$f_{{\tt 6A}}$ & $R_z(\pi)I_z(\frac{\pi}{2})S_z(\frac{\pi}{2})J_{12}(\frac{\pi}{2})$ &
$f_{{\tt 74}}$ & $I_z(-\frac{\pi}{2})R_z(\frac{\pi}{2})J_{12}(\frac{\pi}{2})J_{23}(\frac{\pi}{2})$ \\ \hline
\multicolumn{2}{c|}{Type-II} &
\multicolumn{2}{c|}{Type-III} &
\multicolumn{2}{c}{Type-IV} \\ \hline
$f_{{\tt 1E}}$ & $I_z(\pi)S_z(-\frac{\pi}{2})R_z(-\frac{\pi}{2})J_{23}(\frac{\pi}{2})$ &
$f_{{\tt 4E}}$ & $I_z(\pi)S_z(-\frac{\pi}{2})J_{23}(\frac{\pi}{2})J_{13}(\frac{\pi}{2})$ &
$f_{{\tt 17}}$ & $S_z(\pi)J_{12}(\frac{\pi}{2})J_{23}(\frac{\pi}{2})J_{13}(-\frac{\pi}{2})$ \\ \hline
$f_{{\tt 2D}}$ & $I_z(\pi)S_z(\frac{\pi}{2})R_z(-\frac{\pi}{2})J_{23}(\frac{\pi}{2})$ &
$f_{{\tt 13}}$ & $I_z(\frac{\pi}{2})S_z(\frac{\pi}{2})J_{23}(-\frac{\pi}{2})J_{13}(\frac{\pi}{2})$ &
$f_{{\tt 1B}}$ & $S_z(\pi)J_{12}(\frac{\pi}{2})J_{23}(-\frac{\pi}{2})J_{13}(\frac{\pi}{2})$ \\ \hline
$f_{{\tt 3B}}$ & $I_z(\pi)S_z(-\frac{\pi}{2})R_z(\frac{\pi}{2})J_{23}(\frac{\pi}{2})$ &
$f_{{\tt 27}}$ & $I_z(\frac{\pi}{2})S_z(\frac{\pi}{2})J_{23}(\frac{\pi}{2})J_{13}(-\frac{\pi}{2})$ &
$f_{{\tt 4D}}$ & $S_z(\pi)J_{12}(\frac{\pi}{2})J_{23}(\frac{\pi}{2})J_{13}(\frac{\pi}{2})$  \\ \hline
$f_{{\tt 78}}$ & $I_z(\pi)S_z(\frac{\pi}{2})R_z(\frac{\pi}{2})J_{23}(\frac{\pi}{2})$ &
$f_{{\tt 72}}$ & $I_z(-\frac{\pi}{2})S_z(\frac{\pi}{2})J_{23}(\frac{\pi}{2})J_{13}(\frac{\pi}{2})$ &
$f_{{\tt 71}}$ & $S_z(\pi)J_{12}(-\frac{\pi}{2})J_{23}(\frac{\pi}{2})J_{13}(\frac{\pi}{2})$
\end{tabular}
\end{table}

\end{document}